\documentstyle[preprint,aps]{revtex}

\begin{document}
\draft
\title{Optimal entanglement purifying via entanglement swapping}
\author{Bao-Sen Shi\thanks{%
E-mail address: drshi@ustc.edu.cn}, Yun-Kun Jiang and Guang-Can Guo\thanks{%
E-mail address: gcguo@ustc.edu.cn}}
\address{Laboratory of Quantum Communication and Quantum Computation\\
Department of Physics\\
University of Science and Technology of China\\
Hefei, 230026, P. R. China}
\maketitle

\begin{abstract}
It is known that entanglement swapping can be used to realize entanglement
purifying. By this way, two particles belong to different non-maximally
entangled pairs can be projected probabilisticly to a maximally entangled
state or to a less entangled state. In this report, we show, when the less
entangled state is obtained, if a unitary transformation is introduced
locally, then a maximally entangled state can be obtained probabilisticly
from this less entangled state. The total successful probability of our
scheme is equal to the entanglement of a single pair purification (if two
original pairs are in the same non-maximally entangled states) or to the
smaller entanglement of a single pair purification of these two pairs ( if
two original pairs are not in the same non-maximally entangled states). The
advantage of our scheme is no continuous indefinite iterative procedure is
needed to achieve optimal purifying.
\end{abstract}


Entanglement is at the source of a number of pure quantum phenomena, such as
the correlations violating Bell's inequalities [1], quantum key distribution
[2], quantum teleportation [3], Greanberger-Horne-Zeilinger [GHZ]
correlations [4], and various other nonclassical interference phenomena [5].
Polarization entangled photons have been used to demonstrate both dense
coding [6] and teleportation [7] in the laboratory. Teleportation has also
been realized using path-entangled photons [8] and entangled electromagnetic
field modes [9]. In order to realize these schemes, the entanglement between
distant particles should be set up. One of possible way is entanglement
swapping [10], which has been demonstrated experimentally [11]. Recently,
Bose {\it et al} [12] showed that entanglement swapping can be used to
realize entanglement purifying. In their scheme, if an ensemble of two
photon pair is given, in which all pairs are in the same non-maximally
entangled states, then two photons belong to different photon pairs can be
projected probabilisticly into a maximally entangled Bell state or a less
entangled state by their scheme. If one continues this process
indenfinitely, in the limit of an infinite sequence, the final ensemble
generated will comprise of a certain fraction of Bell pairs and a certain
fraction of completely disentangled pairs. This fraction of Bell pairs
should be equal to twice the modulus square of the Schmidt coefficient of
small magnitude of original pair. In this report, we show that if a unitary
transformation is used followingly when a less entangled state is obtained
by entanglement swapping, then, a maximally entangled Bell state can be
obtained probabilisticly from this less entangled state. The maximum
probability with which a Bell state can be obtained by our scheme is two the
modulus square of the Schmidt coefficient of small magnitude, which means
our scheme is optimal. One advantage of our scheme is no continuous
indenfinite iterative procedure is needed. Furthermore, if two particle
pairs are not the same type of entangled state, two particles belong to
different pairs can also be projected into a maximally entangled state with
certain probability by the same way. This probability is equal to the
smaller entanglement of single pair purification of these two pairs, which
also means our scheme is optimal.

Let pairs of particles (1, 2) and (3, 4) be in the following entangled
states:

\begin{equation}
\left| \Phi \right\rangle _{12}=\alpha \left| 00\right\rangle _{12}+\beta
\left| 11\right\rangle _{12},
\end{equation}

\begin{equation}
\left| \Phi \right\rangle _{34}=\alpha \left| 00\right\rangle _{34}+\beta
\left| 11\right\rangle _{34},
\end{equation}
where, $\left| \alpha \right| >\left| \beta \right| $, and $\left| \alpha
\right| ^2+\left| \beta \right| ^2=1$. Suppose that the particle pair (1, 2)
and the particle 3 belong to Ailce and the particle 4 belongs to Bob. If a
Bell state measurement on particles 2 and 3 is operated by Alice, then the
particles 1 and 4 will be projected into one of the following states

\begin{equation}
\left\langle \Phi ^{\pm }\right| _{23}\left. \Phi \right\rangle _{12}\otimes
\left| \Phi \right\rangle _{34}=\frac{\alpha ^2}{\sqrt{2}}\left|
00\right\rangle _{14}\pm \frac{\beta ^2}{\sqrt{2}}\left| 11\right\rangle
_{14},
\end{equation}

\begin{equation}
\left\langle \Psi ^{\pm }\right| _{23}\left. \Phi \right\rangle _{12}\otimes
\left| \Phi \right\rangle _{34}=\alpha \beta [\frac 1{\sqrt{2}}(\left|
01\right\rangle _{14}\pm \left| 10\right\rangle _{14})].
\end{equation}
Where $\left| \Phi ^{\pm }\right\rangle _{23}=\frac 1{\sqrt{2}}(\left|
00\right\rangle _{23}\pm \left| 11\right\rangle _{23})$ and $\left| \Psi
^{\pm }\right\rangle _{23}=\frac 1{\sqrt{2}}(\left| 01\right\rangle _{23}\pm
\left| 10\right\rangle _{23})$. The particles 1 and 4 will be projected into
a less entangled state with probability $\frac{\alpha ^4+\beta ^4}2$. In
order to get optimal entanglement purifying, a unitary transformation in
Alice' s side ( or in Bob's side, for example, in Alice's side) followes
when a less entangled state is obtained. To carry out this evolution, an
auxiliary qubit with the original state $\left| 0\right\rangle _a$ is
introduced by Alice. Under the basis \{$\left| 0\right\rangle _1\left|
0\right\rangle _a$, $\left| 1\right\rangle _1\left| 0\right\rangle _a$, $%
\left| 0\right\rangle _1\left| 1\right\rangle _a$, $\left| 1\right\rangle
_1\left| 1\right\rangle _a\}$, this unitary transformation can be written as

\begin{equation}
\left[ 
\begin{array}{cccc}
\frac{\beta ^2}{\alpha ^2} & 0 & \sqrt{1-\frac{\beta ^4}{\alpha ^4}} & 0 \\ 
0 & 1 & 0 & 0 \\ 
0 & 0 & 0 & -1 \\ 
\sqrt{1-\frac{\beta ^4}{\alpha ^4}} & 0 & -\frac{\beta ^2}{\alpha ^2} & 0
\end{array}
\right] .
\end{equation}
This transformation will transform Eq. (3) to the follow state

\begin{equation}
\beta ^2[\frac 1{\sqrt{2}}(\left| 00\right\rangle _{14}\pm \left|
11\right\rangle _{14})]\left| 0\right\rangle _a+\frac{\alpha ^2}{\sqrt{2}}%
\sqrt{1-\frac{\beta ^4}{\alpha ^4}}\left| 1\right\rangle _1\left|
0\right\rangle _4\left| 1\right\rangle _a.
\end{equation}
A measurement to the auxiliary particle follows. If the result is $\left|
0\right\rangle _a$, then the particles 1 and 4 will be projected a maximally
Bell state with probability $\beta ^4$ . If the result of the measurement is 
$\left| 1\right\rangle _a$, the particles 1 and 4 are completely
disentangled.

The maximally probability with which a Bell state can be obtained by
purifying a single entangled pair is $2\beta ^2$, which is equal to the
maximum probability with which a Bell state can be obtained, so our scheme
is optimal.

Next, we proceed to consider the case when particle pairs (1, 2) and (3, 4)
are not in the same type of entangled state. Suppose the particles 1 and 2
are in the entangled state $\left| \Phi \right\rangle _{12}$ and the
particles 3 and 4 are in another entangled state $\left| \Phi \right\rangle
_{34}$ which are the following respectively

\begin{equation}
\left| \Phi \right\rangle _{12}=\alpha \left| 00\right\rangle _{12}+\beta
\left| 11\right\rangle _{12}
\end{equation}
and

\begin{equation}
\left| \Phi \right\rangle _{34}=a\left| 00\right\rangle _{34}+b\left|
11\right\rangle _{34}.
\end{equation}
Where $\left| a\right| >\left| b\right| ,$ $\left| a\right| ^2+\left|
b\right| ^2=1.$ Suppose that the particles 1, 2 and 3 belong to Alice and
the particle 4 belongs to Bob. If Alice make a Bell state measurement on
particles 2 and 3, then the particles 1 and 4 will be projected into one of
the follow states

\begin{equation}
\left\langle \Phi ^{\pm }\right| _{23}\left. \Phi \right\rangle _{12}\otimes
\left| \Phi \right\rangle _{34}=\frac{\alpha a}{\sqrt{2}}\left|
00\right\rangle _{14}\pm \frac{\beta b}{\sqrt{2}}\left| 11\right\rangle
_{14},
\end{equation}

\begin{equation}
\left\langle \Psi ^{\pm }\right| _{23}\left. \Phi \right\rangle _{12}\otimes
\left| \Phi \right\rangle _{34}=\frac{\alpha b}{\sqrt{2}}\left|
01\right\rangle _{14}\pm \frac{\beta a}{\sqrt{2}}\left| 10\right\rangle
_{14}.
\end{equation}
If Eq.(9) is obtained, a unitary transformation which is made on the
particle 1 and an auxiliary qubit with the original state $\left|
0\right\rangle _a$ is introduced by Alice (or on the particles 4 and
auxiliary qubit $\left| 0\right\rangle _a$ by Bob). Under the basis \{$%
\left| 0\right\rangle _1\left| 0\right\rangle _a$, $\left| 1\right\rangle
_1\left| 0\right\rangle _a$, $\left| 0\right\rangle _1\left| 1\right\rangle
_a$, $\left| 1\right\rangle _1\left| 1\right\rangle _a\}$, this unitary
transformation is

\begin{equation}
\left[ 
\begin{array}{cccc}
\frac{\beta b}{\alpha a} & 0 & \sqrt{1-\frac{\beta ^2b^2}{\alpha ^2a^2}} & 0
\\ 
0 & 1 & 0 & 0 \\ 
0 & 0 & 0 & -1 \\ 
\sqrt{1-\frac{\beta ^2b^2}{\alpha ^2a^2}} & 0 & -\frac{\beta b}{\alpha a} & 0
\end{array}
\right] .
\end{equation}
Under this transformation, Eq. (9) will be transformed into the state

\begin{equation}
\beta b[\frac 1{\sqrt{2}}(\left| 00\right\rangle _{14}\pm \left|
11\right\rangle _{14})]\left| 0\right\rangle _a+\frac{\alpha a}{\sqrt{2}}%
\sqrt{1-\frac{\beta ^2b^2}{\alpha ^2a^2}}\left| 1\right\rangle _1\left|
0\right\rangle _4\left| 1\right\rangle _a.
\end{equation}
A measurement on the auxiliary particle follows. If the result is $\left|
0\right\rangle _a$, the particles 1 and 4 will be projected into a maximally
entangled state with probability $\beta ^2b^2$. If the result is $\left|
1\right\rangle _a$, the particles 1 and 4 are completely disentangled.

If Eq. (10) is obtained, two different cases should be considered:

1: $\left| \alpha b\right| >\left| a\beta \right| $

In this case, the unitary transformation on the particles 1 and auxiliary
qubit is

\begin{equation}
\left[ 
\begin{array}{cccc}
\frac{\beta a}{\alpha b} & 0 & \sqrt{1-\frac{\beta ^2a^2}{\alpha ^2b^2}} & 0
\\ 
0 & 1 & 0 & 0 \\ 
0 & 0 & 0 & -1 \\ 
\sqrt{1-\frac{\beta ^2a^2}{\alpha ^2b^2}} & 0 & -\frac{\beta a}{\alpha b} & 0
\end{array}
\right] .
\end{equation}
By the same procedure, the particles 1 and 4 will be projected into a
maximally entangled state with probability $a^2\beta ^2$.

2: $\left| \alpha b\right| <\left| a\beta \right| $

In this case, the probability of obtaining a maximally entangled is $\alpha
^2b^2$. The unitary transformation is on the particles 1 and auxiliary qubit
is

\begin{equation}
\left[ 
\begin{array}{cccc}
\frac{\alpha b}{a\beta } & 0 & \sqrt{1-\frac{\alpha ^2b^2}{a^2\beta ^2}} & 0
\\ 
0 & 1 & 0 & 0 \\ 
0 & 0 & 0 & -1 \\ 
\sqrt{1-\frac{\alpha ^2b^2}{a^2\beta ^2}} & 0 & -\frac{\alpha b}{a\beta } & 0
\end{array}
\right] .
\end{equation}
The maximally probability of obtained a maximally entangled state is $2\beta
^2$ or $2b^2$. The first means that entanglement of single pair purification
of the $\left| \Phi \right\rangle _{12}$ is less than that of the state $%
\left| \Phi \right\rangle _{34}$. The second means that the entanglement of
single pair purification of $\left| \Phi \right\rangle _{34}$ is less than
that of the state $\left| \Phi \right\rangle _{12}$. This probability is
equal to the smaller entanglement of single pair purification of these two
pairs before entanglement swapping. Obviously, our scheme is optimal.

In conclusion . In the Ref. [12], an iterative procedure indefinitely is
needed in order to achieve the optimal entanglement purifying. In our
scheme, when a less entangled state is obtained during the entanglement
purifying, if a unitary transformation is introduced locally, a maximally
entangled state can be obtained with certain probability. The successful
probability of our scheme is equal to the entanglement of a single pair
purification if two original pairs are in the same non-maximally entangled
states, or to the smaller entanglement of a single pair purification of
these two pairs if they are not in the same non-maximally entangled states,
which means our scheme is optimal. No continuous indefinite iterative
procedure is needed, which makes our scheme implementable easily in practice.

This subject is supported by the National Natural Science Foundation  of
China (Grant No. 69907005).

\end{document}